\documentclass[aps,twocolumn,showpacs,amsmath,amssymb,pra,reprint]{revtex4-1}

\usepackage{graphicx}%
\usepackage{dcolumn}
\usepackage{bm}
\usepackage{braket}
\begin{document}

\title{High-harmonic generation in $\alpha$-quartz by the electron-hole recombination}
\author{T. Otobe}
\affiliation{Kansai Photon Science Institute, National Institutes for Quantum and Radiological Science and Technology (QST), Kyoto 619-0215, Japan}
\begin{abstract}
A  calculation of the high-harmonic generation (HHG) in $\alpha$-quartz using the time-dependent density functional theory is reported.
The inter-band process is attributed to the dominant in HHG above the band gap.
The photon energy is set to 1.55 eV, and the cutoff energy of the plateau region is found to be limited at the 19th harmonic (30 eV).
The  dependence of the HHG efficiency at the cutoff energy region on laser intensity is consistent with that of the hole density in the lowest-lying valence band. 
Numerical results indicate that electron-hole recombination plays a crucial role in HHG in $\alpha$-quartz.
It is found that a 200 attosecond pulse train is produced using HHG around the plateau cutoff energy.
\end{abstract}
\maketitle
\section{introduction}
In the last four decades, the highly nonlinear interaction between intense ultrafast laser pulses and various materials has attracted great interest.
In particular, high-harmonic generation (HHG) is one of the most important phenomena in laser-matter interactions.
HHG from atoms and molecules enables us attosecond light-pulse generation and reveals the tomography of  molecular orbitals \cite{atto01,Itatani04, Corkum07}.
In general, HHG from atoms and molecules is understood to follow a three-step model: that is, the rescattering process of the emitted electron
with the parent ion \cite{Corkum93}.

Recently, HHG from dielectrics has been reported \cite{Ghimire11, Schubert14, Luu15, Zaks12, Huber15, Langer15,Corkum15,Georges16}.
Because a bulk crystal is a semi-infinite system, the physical process of HHG in a dielectrics should be different from that in  the   gas phase. 
Some theoretical models to describe HHG from bulk crystals have been proposed,   including Bloch oscillation \cite{Ghimire11, Schubert14, Luu15,Hawkins13,Hawkins15}, electron-hole recombination\cite{Zaks12, Huber15,  Langer15, Corkum15, Vampa14, Vampa15,Georges16}, and  Wannier-Stark localization \cite{Higuchi14}).
However, further experimental and theoretical investigation is needed to clarify which effect is the dominant process.
 
 Previously, we  successfully  applied the time-dependent density-functional theory (TDDFT) \cite{Runge84} to laser-matter interactions\cite{Otobe08, Shino10, yabana12, Sato14, Sato15,Otobe16}.
 We also solved the time-dependent Kohn-Sham (TDKS) equation, which is the fundamental equation of the TDDFT,  
using the real-time and -space method \cite{Bertsch00}.
 Because  our method does not require preparation of the excited states, highly nonlinear processes can be described with relatively low computational cost.
 
In this paper, we report a density-functional calculation using  real-time TDDFT of HHG in $\alpha$-quartz under an intense laser field.
We chose $\alpha$-quartz as a suitable  target for HHG because its  band structure  is very simple. 
The effective mass of the  valence band  of  $\alpha$-quartz is very heavy ($5\sim 10 m$), and  only the conduction band around the $\Gamma$-point has light effective mass (0.3$m$, where $m$ is the electron mass) \cite{Chelikowsky77}.
Therefore, the comparison between the HHG spectrum and band structure of $\alpha$-quartz is straightforward.

\section{real-time TDDFT calculation}
In real-time TDDFT, we described the electron dynamics in a unit  
cell of a crystalline solid under a spatially uniform electric field $E(t)$. 
Treating the field with a vector potential, we obtained 
\begin{equation}
\vec A(t)=-c\int^t dt' \vec E(t'). 
\end{equation}
We assumed the laser was linearly polarized,  and the polarization direction was parallel to the optical axis of $\alpha$-quartz.
The electron dynamics were described by the following 
TDKS equation in atomic units (a.u.): 
\begin{equation} 
i \frac{\partial}{\partial t} \psi_i(\vec{r},t)= 
\left[ \frac{1}{2m} \left( \vec p + \frac{e}{c} \vec A(t) \right)^2 
+ V(\vec r,t) \right] \psi_i (\vec r,t), 
\label{TDKS} 
\end{equation} 
where $V(\vec r,t)$ is composed of the 
electron-ion, Hartree, and exchange-correlation potentials.  
We use a norm-conserving pseudopotential for the electron-ion  
potential 
\cite{TM91,Kleinman82}. 
In this work we used a modified Becke-Johnson (mBJ) exchange potential\cite{mBJ} as given by Eqs. (2)-(4) in Ref. \cite{mBJ2} with a  local-density approximation (LDA) correlation potential \cite{LDA} under adiabatic approximation.

The rectangular unit cell containing six silicon atoms and 12 oxygen atoms was discretized into Cartesian grids of $26\times 44\times 38$.
 The point group of the crystal structure is P3(2)21.
 The $k$ space was also discretized into $ 8^3$ grid points. 
 Time evolution was computed using a fourth-order Taylor expansion of the operator.
 We use a time step of 0.02 a.u. The number of time steps was typically 40,000.
 
\section{ground state and linear response of $\alpha$-quartz}

\begin{figure}
\includegraphics[width=90mm]{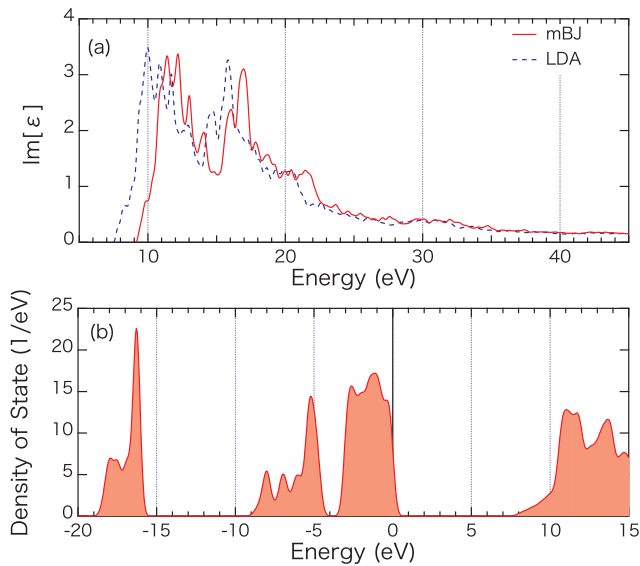}
\caption{\label{fig:Fig0}(a)Imaginary part of the dielectric function (${\rm Im}\epsilon$) with LDA (blue dashed line) and mBJ (red solid line) potentials.
(b) DoS of the $\alpha$-quartz with mBJ potential.} 
\end{figure}

  Figure~\ref{fig:Fig0} (a) shows the imaginary part of the dielectric function (${\rm Im}\epsilon$) calculated by real-time approach \cite{yabana12}
   with LDA (blue dashed line) and mBJ (red solid line) potentials.
 mBJ potential gives an optical band gap of 9.2 eV for $\alpha$-quartz agreeing well the experimental value of 9-10 eV.
 Figure~\ref{fig:Fig0} (b) shows the density of states (DoS) as the function of energy from the top of the valence band, which shows the lower band gap ($\sim 8$ eV).

\begin{figure}
\includegraphics[width=80mm]{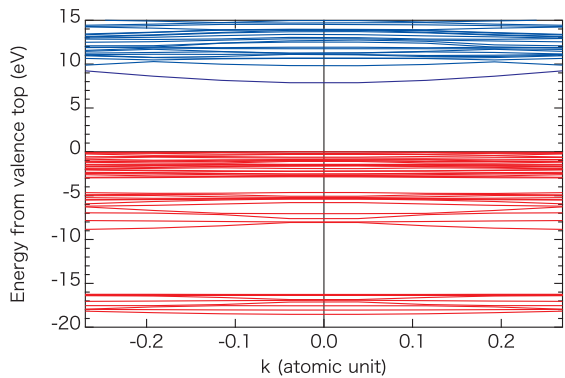}
\caption{\label{fig:Sup_Fig1}  Band map along the $\vec{k}=k(0,0,1)$ line in Cartesian coordinate.}
\end{figure}

  Although our calculation employ the primitive cell, the band map of the ground state may give important information.
 Figure ~\ref{fig:Sup_Fig1} show the band map along $\vec{k}=k(0,0,1)$ line.
 Blue (red) lines  presents the conduction bands (valence bands). 
 Valence top has very flat structure and heavy effective mass.
 Therefore the difference between the optical gap (Fig.~\ref{fig:Fig0} (a)) and energy gap in DoS (Fig.~\ref{fig:Fig0} (b)) 
 is attributed to the dynamical effect in TDDFT.
 
\section{High-harmonic generation in $\alpha$-quartz}
\subsection{HHG under few cycle  intense laser}
The laser electric field $E(t)$ was assumed to be,
\begin{equation}
E(t)=
\begin{cases}
E_0 \sin^2\left(\pi \frac{t}{T_p}\right)\cos(\omega_0 t+\phi) & 0<t<T_p \\
0& T_p<t< T_e, \label{eq:field}
\end{cases}
\end{equation} 
where $E_0$ is the maximum electric field amplitude and  $\omega_0$ is the laser frequency ($\omega_0=1.55$ eV).
 $E_0$ is related to the incident laser field ($E_{\rm in}$) by $E_0=2/(1+\sqrt{\varepsilon})E_{\rm in}$, where $\varepsilon$ is the dielectric function at $\omega_0$ \cite{yabana12}.
The pulse length $T_p$ was set to be six optical cycles (16.2 fs), and  the computation was terminated at $t=T_e$ (19.4 fs).

An important output of the calculation was the
average electric current density as a function of time, $J(t)$, which is given by 
\begin{equation} 
J(t) = -\frac{e}{mV}  \int_{V} d\vec r \sum_i 
{\rm Re} \psi_i^* \left( \vec p + \frac{e}{c}\vec A(t) \right) \psi_i + J_{ion}(t), 
 \label{current} 
\end{equation} 
where $V$ is the volume of the unit cell.  
$J_{ion}(t)$ is the current produced by the pseudopotential \cite{Bertsch00, Sato15}. 
The spectrum of light inside a material can be calculated by the Fourier transformation
of the electron current,
\begin{equation}
I(\omega )=\left|\int_{0}^{T_e}J(t)\exp(-i\omega t) dt\right|^2.
\end{equation}

\begin{figure}
\includegraphics[width=90mm]{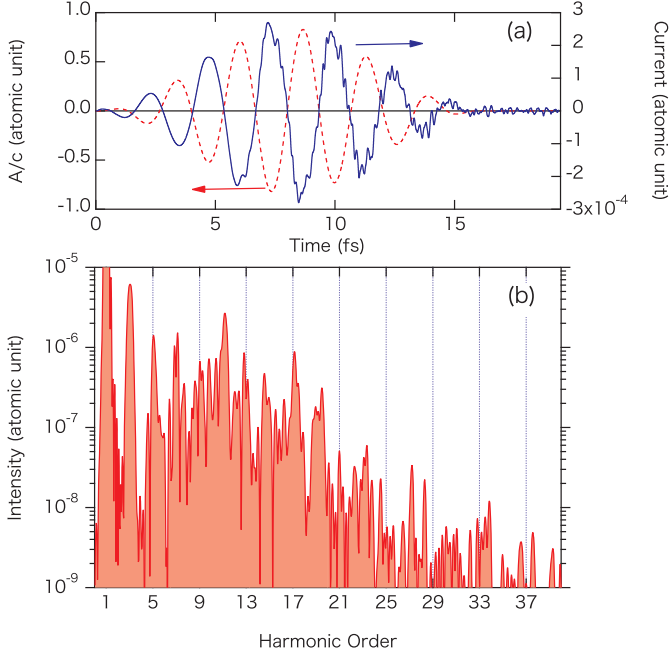}
\caption{\label{fig:Fig1}(a) Applied vector potential (red dashed line) and  induced electronic current (blue solid line).
(b) HHG spectrum calculated from the Fourier transformation of the  electronic current. } 
\end{figure}
 
We present the result calculated at the laser intensity of $8\times10^{13}$ W/cm$^2$ as Fig.~\ref{fig:Fig1}.
The laser intensity is related to the applied electric field by $I_0=cE_0^2/8\pi$.
$I_0$ in the bulk is related to incident intensity $I_{ \rm in}$ by $I_{\rm in}=1.327 I_0$ \cite{yabana12}.
Therefore,  $8\times10^{13}$ W/cm$^2$ is consistent with $I_{\rm in}=1.06\times10^{14}$ W/cm$^2$.
The red dashed line in Fig.~\ref{fig:Fig1} (a) indicates the applied vector potential ($A(t)/c$) with  phase $\phi=0$ in Eq.~(\ref{eq:field}), 
and the blue solid line is  the induced current (J(t)).
The intensity of HHG, $I(\omega )$, is shown in Fig.~\ref{fig:Fig1}(b).
The HHG spectrum is characterized by two relatively intense peaks at 11th harmonic and 17th harmonic.

\begin{figure}
\includegraphics[width=90mm]{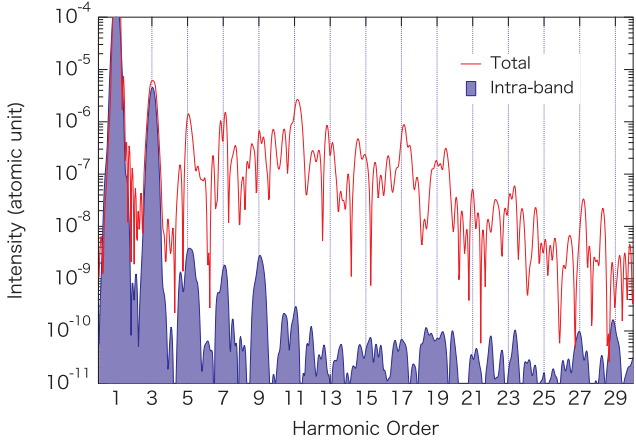}
\caption{\label{Inter_Intra} HHG spectrum calculated by intra-band interaction model (filled blue line) is compared with the spectrum including all processes (red solid line). } 
\end{figure}
Previous works on HHG in solids discussed the contribution of the inter-band\cite{Zaks12, Huber15,  Langer15, Corkum15, Vampa14, Vampa15,Georges16, Higuchi14} and intra-band\cite{Ghimire11, Schubert14, Luu15,Hawkins13,Hawkins15} processes.
One possible way to elucidate the physical process is expanding the electron dynamics into inter- and intra-band processes. 
The time-dependent wavefunctions can be projected to the eigenstates of the time-dependent hamiltonian ($\Phi_i$), 
\begin{equation}
\psi_{i}(t)\approx \sum_{i'} \langle \Phi_{i'} \vert \psi_{i}(t) \rangle \Phi_{i'} \equiv  \sum_{i'}C_{i'i}(t) \Phi_{i'},
\end{equation}
where the band index $i'$ runs valence and conduction bands.
If we can prepare the complete set of $\Phi_i$, the current $J(t)$ is expanded into intra-band 
\begin{eqnarray} 
J_{intra}(t) &=& -\frac{e}{mV}  \int_{V} d\vec r \sum_{i'i}|C_{i'i}(t)|^2 
{\rm Re} \Phi_{i'}^* \left( \vec p + \frac{e}{c}\vec A(t) \right) \Phi_{i'} \nonumber\\
&&+ J_{ion,intra}(t), 
 \label{current_int} 
\end{eqnarray}
and inter-band
\begin{eqnarray} 
J_{inter}(t) &=& -\frac{e}{mV}  \int_{V} d\vec r \sum_{i'\ne i" i}C_{i'i}^*(t)C_{i"i}(t)\nonumber\\
&\times&{\rm Re} \Phi_{i'}^* \left( \vec p + \frac{e}{c}\vec A(t) \right) \Phi_{i''} + J_{ion,inter}(t), 
 \label{current_int} 
\end{eqnarray}
components exactry.
 Here $J_{ion,intra(inter)}(t)$ is intra- (inter-)comtribution from pseudopotential.
The intra-band component is accessible because it dose not depend on the  phase of $\Phi_i$ and $C_{i'i}(t)$.

Figure~\ref{Inter_Intra} shows the HHG from $J(t)$ and $J_{intra}(t)$  at the laser intensity of $8\times10^{13}$ W/cm$^2$.
We assumed 10 conduction bands in this calculation for intra-band process, 
and calculated the $\Phi_i$ and $C_{i'i}$ each 50 time steps, which corresponds to one atomic unit in time.

The HHG spectrum by intra-band process shows relatively good agreement with full calculation at 3rd harmonic.
However it drop off at 5th harmonic and the signal decreases to numerical error level above 11th harmonic.
Therefore, contribution of the intra-band process can be considered as minor effect and the inter-band process is dominant.

\begin{figure}
\includegraphics[width=90mm]{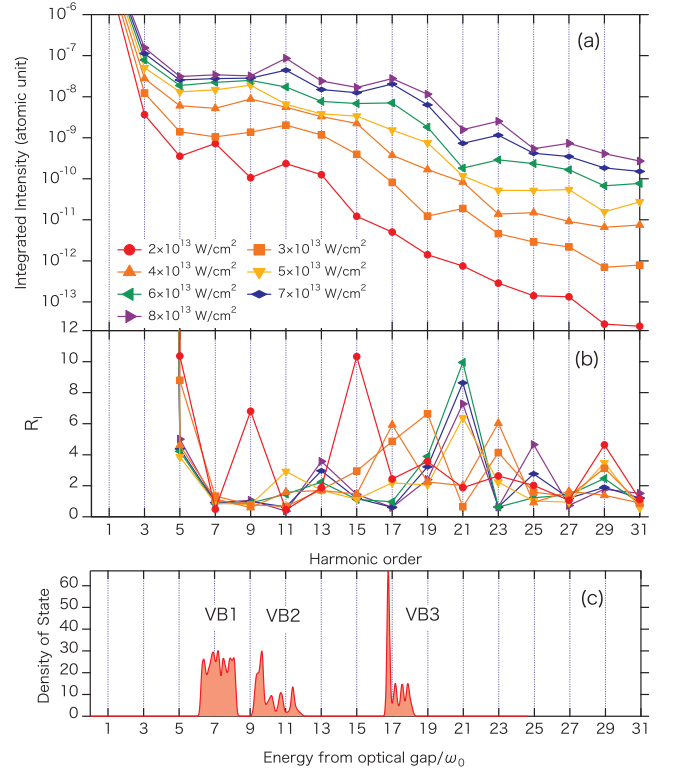}
\caption{\label{fig:Fig2} Dependence of the HHG on laser intensity. (a) Integrated HHG spectrum calculated at 2-8 $\times 10^2$ W/cm$^2$.
 (b) Relative intensity ($R_l$) defined by Eq.~(\ref{Rel_Int}).
(c) DoS of the valence bands. The horizontal axis is the absolute energy of the optical gap (9.2 eV) scaled by  $\omega_0$. }
\end{figure}

\subsection{Intensity dependence}
In previous subsection, we confirm that the inter-band process is dominant in HHG.
In this subsection, we would like to explore detail of HHG.

The laser intensity dependence of the integrated HHG spectra,
\begin{equation}
\tilde{I}_l=\int_{(l-1)\omega_0}^{(l+1)\omega_0} d\omega I(\omega),
\end{equation}
 is presented in Fig.~\ref{fig:Fig2} (a).
 From $I_0=2\times 10^{13}$ to $4\times 10^{13}$ W/cm$^2$, some peaks appear at different harmonic order, 
and the spectrum accesses to flat structure (plateau).
Above $I_0=5\times 10^{13}$ W/cm$^2$, the spectral shape becomes similar. 
 
In most case, the definition of the cutoff energy is difficult because the HHG spectrum in solids dose not drop off clearly like gas phase\cite{Lewenstein}.
 Figure~\ref{fig:Fig2} (b) presents the relative intensity of each harmonics,
\begin{equation}
\label{Rel_Int}
R_l=\frac{\tilde{I}_{l-2}}{\tilde{I}_{l}},
\end{equation}
which presents how the $l$-th order harmonic decreases with respect to $(l-2)$-th order harmonic.
The legend of Fig.~\ref{fig:Fig2} (b) is the same as Fig.~\ref{fig:Fig2} (a).
In this paper, we would like to define the cutoff as the intense single peak in Fig.~\ref{fig:Fig2} (b).
Above $I_0=5\times 10^{13}$ W/cm$^2$, single intense peak can be seen at 21th harmonic.
This result indicates that the cutoff is stay at 19th harmonic above $I_0=5\times 10^{13}$ W/cm$^2$.

The robustness of the cutoff energy is an important feature, because both the Bloch oscillation\cite{Ghimire11, Luu15} and  Wannier-Stark localization \cite{Higuchi14} are sensitive to laser intensity. 
In contrast, the  electron-hole recombination model is consistent with our result \cite{Georges16, Vampa14,Vampa15}.
In particular, the experimental and theoretical result reported by Ndabashimiye {\it et al.} \cite{Georges16} suggest that the saturated plateau cutoff corresponds to the recombination  between valence  and conduction bands in a  rare-gas solid.  

In general, the velocity of electrons and holes depends on the inverse of the effective mass.
In the case of $\alpha$-quartz, the reduced mass (or effective mass of electron) has the smallest value around the $\Gamma$-point, which corresponds to the direct band gap as shown in Fig.~\ref{fig:Sup_Fig1}.

The density of states  of the valence band for the initial state  are presented as a function of the absolute energy from the optical band gap in Fig.~\ref{fig:Fig2} (b).
We defined the three valence bands as VB1, VB2, and VB3.
The band energy was scaled by the photon energy, $\omega_0$, to allow comparison with the HHG spectrum.
The plateau cutoff energy coincides with the energy of VB3, and  the characteristic peaks at the 11th and 17th harmonics for $I_0=8\times 10^{13}$ W/cm$^2$ correspond to  VB2 and VB3 respectively.
These results also indicate that HHG in $\alpha$-quartz can be understood on the basis of  the electron-hole recombination process.

We note that the orbital energy in this DFT calculation has less meaning than that of an orbital-dependent theory, such as the Hartree-Fock theory.
In our calculation, the direct band gap calculated from the orbital energies is 7.8 eV, which is smaller than the optical band gap of $\alpha$-quartz of 9.2 eV calculated from a real-time simulation (Fig.~\ref{fig:Sup_Fig1} and Ref. \cite{Sato15}).  
Therefore, the horizontal axis of Fig.~\ref{fig:Fig2}(b) has an uncertainty of 10\% or 1 eV.
This uncertainty  is relatively small and does not affect our conclusions.

\begin{figure}
\includegraphics[width=90mm]{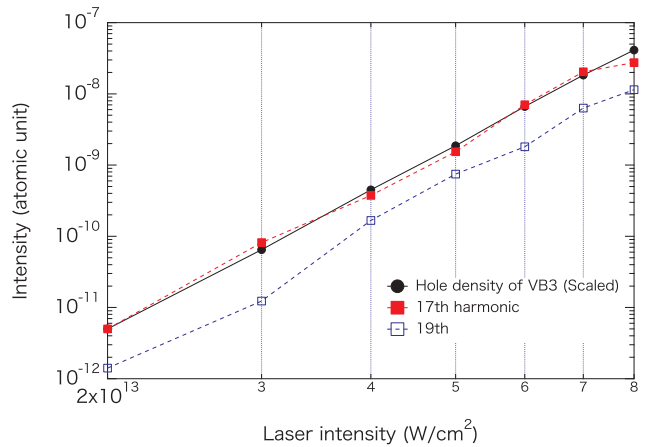}
\caption{\label{fig:Fig3}  Dependence of the integrated intensity of the  17th (red filled squares) and 19th (blue open squares) harmonics on laser intensity. 
 The hole density in VB3 is indicated by the black solid line with filled circles. The hole density at $2\times 10^{13}$  W/cm$^2$ is scaled to the 17th harmonic.}
\end{figure}

The relation between the HHG spectrum and hole density of each valence band should be consistent with the recombination process.
In previous works, we defined the hole density as the projection of the time-dependent wavefunction at $t=T_e$ to the initial state given by, 
\begin{equation}
n_{\rm ex} = \frac{1}{V}\sum_{ii' =occ} \left( \delta_{ii'} -  
\vert \langle \Phi_{i} \vert \psi_{i'}(t=T_e) \rangle \vert^2 \right),
\end{equation}
where $i$ and $i'$ are the band indices of orbitals for the initial and time-dependent states, and $\Phi_i$ is the wavefunction of the initial state, $i$ \cite{Otobe08}.
Similarly, the hole density for VB3 is defined as,
\begin{equation}
n_{\rm ex}^{{\rm VB3}} = \frac{1}{V}\sum_{i={\rm VB3} }\sum_{i' =occ} \left( \delta_{ii'} -  
\vert \langle \Phi_{i} \vert \psi_{i'}(t=T_e) \rangle \vert^2 \right).
\end{equation}
Figure \ref{fig:Fig3} shows the laser-intensity dependence of  17th and  19th harmonics, together with the $n_{\rm ex}^{{\rm VB3}}$.
We scaled the hole density at $I_0=2\times10^{13}$ W/cm$^2$ to obtain the intensity dependence.
The power laws of hole density and harmonic intensity agree very well.
This result strongly indicates that the 17th harmonic is caused by the electron-hole recombination in VB3.
We also present the results for the  19th harmonic, which is the neighbor of the 17th harmonic in Fig.~\ref{fig:Fig3}.
Because the 17th and 19th harmonics show the same power law, they both correspond to the emission caused by the electron-hole recombination in VB3. 
It should be noted that the Keldysh parameter \cite{Kel65} for VB3 is around one in the present laser intensities.
Therefore, the laser-intensity dependence of the hole density is  not scaled to the  usual $I_0^a$ power law where $a$ is the photon number.

The laser-intensity dependence of lower-order harmonics shows fluctuation and saturation,  as illustrated in  Fig~\ref{fig:Fig2}(a).  
The HHG from different paths interfere, which decrease its overall intensity \cite{Hawkins13,Hawkins15}.
At higher intensity, electron excitation to higher-lying conduction bands becomes substantial.
Therefore, the emission from various quantum paths contributes to HHG, which modulates the HHG intensity.
For higher harmonics concerning the lowest-lying valence band, the laser intensity is not sufficient to cause such interference with the HHG from VB2  because the  energy difference is large ($\sim 6$ eV).
According to our previous work, at extremely high laser intensities, the $\alpha$-quartz is excited so strongly that the laser-plasma interaction becomes a major process when a near-infrared laser is used \cite{Sato15,Otobe12}.
In other words, HHG around the plateau cutoff is emitted from same quantum path and maintains good coherence in our calculation.

\subsection{Time evolution of HHG}
\begin{figure}
\includegraphics[width=90mm]{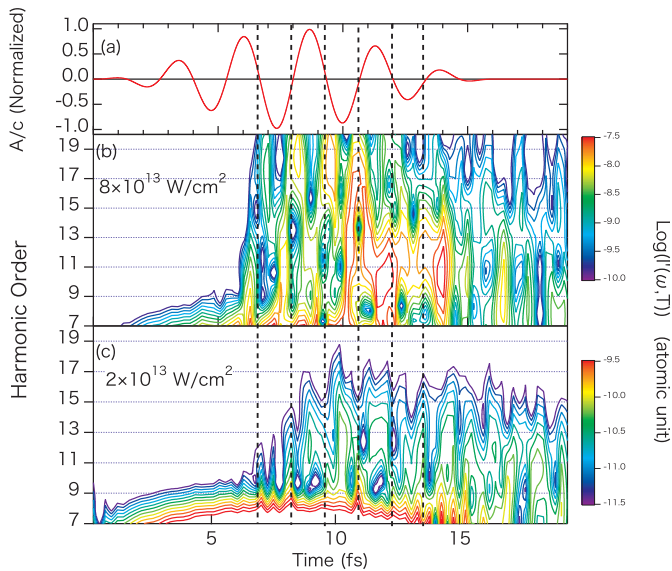}
\caption{\label{fig:Fig4} HHG intensity as functions of time and frequency ($I'(\omega,T)$). (a) Time evolution of the normalized vector potential. 
 $I'(\omega,T)$ for laser intensities of (b) $8\times 10^{13}$ and (c) $2\times 10^{13}$ W/cm$^2$ are presented respectively. }
\end{figure}
The time evolution of HHG emission ($I'(\omega,T)$) calculated from the time-gated Fourier transformation;
\begin{equation}
\label{eq:G_FT}
I'(\omega,T)=\left| \int_0^{T_e} J(t)\exp(-i\omega t)\exp\left(-\frac{(t-T)^2}{\eta^2}\right) dt\right|^2,
\end{equation}
is shown in Fig.~\ref{fig:Fig4}(b) and (c). 
Here, we define the parameter $\eta$ as 0.24 fs (ten a.u. ) in Eq.~(\ref{eq:G_FT}).
Figure~\ref{fig:Fig4}(a) is the normalized $A(t)/c$, and the vertical dashed lines represent the minimum of $A(t)$, which corresponds to the maximum of the electric field.
At an intensity of $2\times 10^{13}$ W/cm$^2$, the 9-13th and 15-19th harmonics show almost the same time dependence.
In contrast, at a higher intensity of  $8\times 10^{13}$ W/cm$^2$,  the relative phase of each HHG emission with respect to the laser field shifts between the 9-13th and 15-19th harmonics.
 In particular, the time dependence of 15-19th harmonics is  relatively strong compared with that of the 9-13th harmonics.
Because the 9-13th and 17-19th harmonics can be attributed to VB2 and VB3, respectively, from Fig.~\ref{fig:Fig2},
the relative phase shift also indicates that these two energy region have different quantum paths.
The 15th harmonics shows intermediate time-evolution between those of the 9-13th and 17-19th harmonics. 
\begin{figure}
\includegraphics[width=90mm]{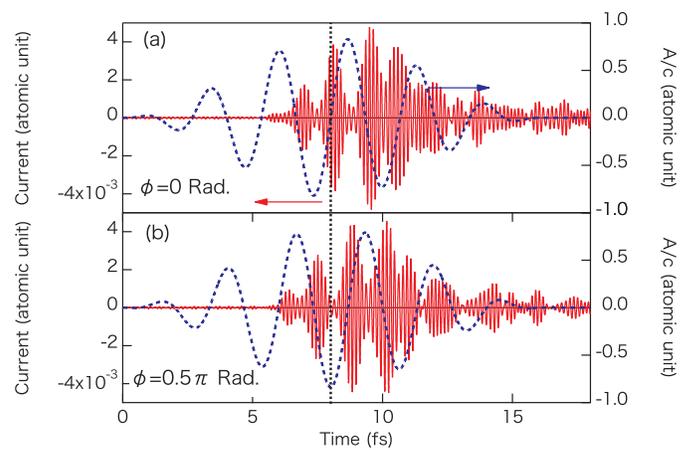}
\caption{\label{fig:Fig5} Time evolution of the current, $\tilde{J}_n(t)$ by the $n=$15th, 17th and 19th  harmonics (red solid line).
Blue dashed lines indicate the applied vector potentials. (a) The phase of the applied laser, $\phi$ in Eq. (\ref{eq:field}),  is zero. 
(b) The $\phi=0.5 \pi$. The relative phase of the pulse train to the driving field is independent of $\phi$.}
\end{figure}

Because the  15-19th harmonics have the same time evolution and quantum paths, their summation is expected to be an ultrafast pulse on the  attosecond time scale. 
The time evolution of the current corresponding to the $n$-th harmonics ($\tilde{J}_n(t)$) from the filtered inverse Fourier transformation is given by,
\begin{equation}
\tilde{J}_n(t)=\frac{1}{2\pi}\int_{(n-1)\omega_0}^{(n+1)\omega_0}  \int_{0}^{T_e}dt' d\omega J(t')\exp(-i\omega (t'-t)) .
\end{equation}
Figure \ref{fig:Fig5}(a) depicts the summation of the current for the 15- 19th  harmonics (red solid line)  and the applied vector potential (blue dashed line) for an intensity of $8\times 10^{13}$ W/cm$^2$.
The vertical dotted line indicates the peak of the envelope function of the laser pulse.
The 15-19th  harmonics show a burst at the minimum vector potential (maximum  electric field) and the pulse train has a duration of about 200 attoseconds.

Figure \ref{fig:Fig5}(b) displays the case for the  $\phi=0.5\pi$.
The relative phase of the pulse train with respect to the laser field is unchanged.
In the case of $\phi=0$, the pulse train has a single maximum peak at 9.5 fs.
However, Fig.~\ref{fig:Fig5}(b) showing $\phi=0.5\pi$ contains  double maximum peaks .
Therefore, the intensity of the pulse train constructed from the cutoff region is the controlled by 
the phase of the applied incident laser.

\section{Summary}
In summary, we reported a density-functional calculation of the HHG in $\alpha$-quartz under an intense few-cycle pulse laser.
Our simulation indicates that the inter-band interaction is the dominant process. 
We found that the cutoff of the plateau was limited at the 19th harmonic (30 eV), 
which corresponds to the maximum energy gap between the lowest-lying valence band and the bottom of the conduction band.
The power laws of the hole density and the harmonic intensity  indicate that the HHG emission in an $\alpha$-quartz is attributable to electron-hole recombination. 
HHG around the cutoff energy enabled us to generate a pulse train of 200 attoseconds from the solid target because the electron velocity was high at
the specific $k$ point with a small mass ($\Gamma$-point in $\alpha$-quartz).

\section*{Acknowledgement}
This work was supported by a JSPS KAKENHI (Grants No. 15H03674). 
Numerical calculations were performed on the supercomputer SGI ICE X at 
the Japan Atomic Energy Agency (JAEA).


\begin{thebibliography}{99}
\bibitem{atto01} M. Hentschel, R. Klenberger, Ch. Spielmann, G.A. Reider, N. Milosevic, T. Brabec, U. Heinzmann, M. Drescher, and F. Krausz, Nature {\bf 414}, 509 (2001). 
\bibitem{Itatani04} J. Itatani, J. Levesque, D. Zeidler, H. Niikura, H. Pepin, J. C. Kieffer, P. B. Corkum, and D. M. Villeneuve, Nature {\bf 432}, 867 (2004). 
\bibitem{Corkum07} P. B. Corkum and Ferenc Krausz, Nature Physics {\bf 3}, 381 (2007).
\bibitem{Corkum93} P. B. Corkum, Phys. Rev. Lett. {\bf 71}, 1994 (1993).
\bibitem{Ghimire11} S. Ghimire, A. D. DiChiara, E. Sistrunk, P. Agostini, L. F. DiMauro and D. A. Reis, Nature Physics {\bf 7} 138, (2011).
\bibitem{Schubert14} O. Schubert, M. Hohenleutner, F. Langer, B. Urbanek, C. Lange, U. Huttner, D. Golde, T. Meier, M. Kira, S. W. Koch, and R. Huber, Nat. Photonics {\bf 8}, 119 (2014)
\bibitem{Luu15} T. T. Luu, M. Garg, S. Yu. Kruchinin, A. Moulet, M. Th. Hassan, and E. Goulielmakis, Nature {\bf 521} 498, (2015).
\bibitem{Zaks12} B. Zaks,	R. B. Liu, and M. S. Sherwin, Nature {\bf 483} 580, (2012).
\bibitem{Huber15} M. Hohenleutner, F. Langer, O. Schubert, M. Knorr, U. Huttner, S. W. Koch, M. Kira, and R. Huber, Nature {\bf 523} 572, (2015).
\bibitem{Langer15} F. Langer, M. Hohenleutner, C. P. Schmid, C. Poellmann, P. Nagler, T. Korn, C. Schuller, M. S. Sherwin, U. Huttner, J. T. Steiner, S. W. Koch, M. Kira, and R. Huber, Nature {\bf 533} 225, (2016).
\bibitem{Corkum15} G. Vampa, T. J. Hammond, N. Thire, B. E. Schmidt, F. Legare, C. R. McDonald, T. Brabec, and P. B. Corkum, Nature {\bf 522} 462, (2015).
\bibitem{Georges16} G. Ndabashimiye, S. Ghimire, M. Wu, D. A. Browne, K. J. Schafer, M. B. Gaarde, and D. A. Reis, Nature {\bf 534}, 520 (2016).
\bibitem{Hawkins13} P. G. Hawkins and M. Y. Ivanov, Phys. Rev. A {\bf 87} 063842 (2013).
\bibitem{Hawkins15} P. G. Hawkins, M. Y. Ivanov, and V. S. Yakovlev, Phys. Rev. A {\bf 91} 013405 (2015).
\bibitem{Vampa14} G. Vampa, C. R. McDonald, G. Orlando, D. D. Klug, P. B. Corkum, and T. Brabec, Phys. Rev. Lett. {\bf 113} 073901 (2014).
\bibitem{Vampa15} G. Vampa, C. R. McDonald, G. Orlando, P. B. Corkum, and T. Brabec, Phys. Rev. B {\bf 91} 064302 (2015).
\bibitem{Higuchi14} T. Higuchi, M. I. Stockman, and P. Hommelhoff, Phys. Rev. Lett. {\bf 113} 213901, (2014).
\bibitem{Runge84} E. Runge and E. K. U. Gross,  Phys.Rev. Lett. {\bf 52}, 997 (1984). 
\bibitem{Otobe08}  T. Otobe, M. Yamagiwa, J. -I. Iwata, K. Yabana, T. Nakatsukasa, and G. F. Bertsch,  Phys. Rev. B{\bf77}, 165104 (2008).
\bibitem{Shino10} Y. Shinohara, K. Yabana, Y. Kawashita, J.-I. Iwata, T. Otobe , and George F. Bertsch,  Phys. Rev. B {\bf 82}, 155110 (2010). 
\bibitem{yabana12} K. Yabana, T. Sugiyama, Y. Shinohara, T. Otobe, and G. F. Bertsch, Phys. Rev. B  {\bf 85}, 045134 (2012). 
\bibitem{Sato14} S. A. Sato, K. Yabana, Y. Shinohara, T. Otobe, G.F. Bertsch, Phys. Rev. {\bf B89}, 064304 (2014).
\bibitem{Sato15}  S. A. Sato, K. Yabana, Y. Shinohara, T. Otobe,K.-M. Lee, and  G.F. Bertsch, Phys. Rev. B {\bf 92} 205413 (2015).
\bibitem{Otobe16} T. Otobe, Y. Shinohara, S. A. Sato, and K. Yabana, Phys. Rev. B {\bf 93}, 045124 (2016). 
\bibitem{Bertsch00} G.F. Bertsch, J.-I. Iwata, A. Rubio, and K. Yabana, Phys. Rev. B {\bf62} , 7998 (2000). 
\bibitem{Chelikowsky77} J. R. Chelikowsky and M. Schluter, Phys. Rev. B {\bf 15}, 4020 (1977).
\bibitem{TM91} N. Troullier and J.L. Martins, Phys. Rev. {\bf B43}, 1993 (1991). 
\bibitem{Kleinman82} L. Kleinman and D. M. Bylander, Phys. Rev. Lett.  {\bf 48}, 1425 (1982). 
\bibitem{mBJ} A. D. Becke and E. R. Johnson, J. Chem. Phys. {\bf 124} 221101 (2006).
\bibitem{mBJ2} F. Tran and P. Blaha, Phys. Rev. Lett. {\bf 102}, 226401 (2009).
\bibitem{LDA} J. P. Perdew and Y. Wang, Phys. Rev. B {\bf 45}, 13244 (1992).
\bibitem{Lewenstein} M. Lewenstein, Ph. Balcou, M. Yu. Ivanov, Anne L'Huillier, and P. B. Corkum, Phys. Rev. A {\bf 49}, 2117 (1994). 
\bibitem{Kel65} L.V. Keldysh, Sov. Phys. JETP {\bf 20}, 1307 (1965).	
\bibitem{Chin} S. L. Chin, \textit{Advances in Multiphoton Processes and Spectroscopy},  (World Scientific, Singapore 2004), Chapter 3 .
\bibitem{Otobe12} T. Otobe, Appli. Phys. Lett. {\bf 111}, 093112 (2012). 
\end{thebibliography}
\end{document}